\documentclass{article}
\usepackage{spconf,amsmath,graphicx}
\usepackage{epstopdf}
\usepackage{amssymb,amsmath,amsthm}
\usepackage{xifthen}
\usepackage[english]{babel}
\usepackage{graphicx,xcolor}
\graphicspath{{figures/}}

\newcommand{\ing}[1]{\mathsf{#1}}
\newcommand{\Rn}[1]{\ifthenelse{\equal{#1}{}}{\mathbb{R}}{\mathbb{R}^{\ing{#1}}}}
\newcommand{\Cn}[1]{\mathbb{C}^{\ing{#1}}}
\newcommand{\Rset}[2]{\ifthenelse{\equal{#2}{1}}{\in \Rn{\ing{#1}}}{\in \Rn{\ing{#1} \times \ing{#2}}}}
\newcommand{\Cset}[2]{\ifthenelse{\equal{#2}{1}}{\in \Cn{\ing{#1}}}{\in \Cn{\ing{#1} \times \ing{#2}}}}
\newcommand{\vect}[1]{\boldsymbol{\mathbf{\MakeLowercase{#1}}}}
\newcommand{\mtrx}[1]{\boldsymbol{\mathbf{#1}}}

\newcommand{\norm}[2]{\|#1\|_{#2}}

\newcommand{\transp}[1]{#1^{\mathsf{T}}}

\newcommand{\ind}[2]{\ifthenelse{\equal{#2}{}}{\chi_{#1}}{\chi_{#1}\left( #2 \right)}}

\newcommand{\iter}[2]{#1^{\ing{(#2)}}}

\DeclareMathOperator*{\argmax}{argmax\,}

\definecolor{light-gray}{gray}{0.75}

\usepackage{siunitx}

\title{Time Domain Velocity Vector for Retracing the multipath propagation}
\name{J{\'e}r{\^o}me Daniel, Sr{\dj}an Kiti{\'c}} 
\address{Orange Labs, France \thanks{Both authors equally contributed to this work.}}

\begin{document}

%
\maketitle
\begin{abstract}
We propose a conceptually and computationally simple form of sound velocity, which seamlessly exploits active and reactive sound intensity measurements, and, consequently, provides a more transparent view of acoustic multipath. We feel that this representation has a potential both as a valuable tool for directly analyzing sound propagation, as well as being a new spatial feature format for machine learning algorithms in audio and acoustics. As a showcase, we demonstrate that the Direction-of-Arrival (DoA) \emph{and} the range of a sound source could be estimated from the First Order Ambisonics (FOA) recordings, the latter never attempted before, to the best knowledge of the authors.
\end{abstract}

\section{Introduction}
\label{sec:intro}

Sound intensity is one of the most fundamental acoustic quantities, and has found its role in various applications, from coding \cite{pulkki2007spatial}, to soundfield synthesis \cite{daniel1998ambisonics, spors2007extension}, psychoacoustics \cite{merimaa2006analysis,rossing2007springer} and acoustic holography \cite{rossing2007springer}. Theoretically defined as an instantaneous temporal quantity, sound intensity is most often expressed in the frequency domain, assuming the harmonic decomposition of a soundfield \cite{jacobsen1991note}. Its real part is termed the \emph{active intensity}, as it describes the flow of energy originating from a sound source \cite{fahy2000foundations,rossing2007springer}. The behavior of its imaginary part, the \emph{reactive intensity}, is somewhat less evident, as it describes the non-propagative local energy exchange that depends on the sound waves' interference \cite{merimaa2006analysis, daniel2000representation}.

Narrowband active sound intensity has been widely used for sound source localization in Ambisonics, being a computatuonally cheap alternative to beamforming techniques \cite{jarrett20103d, pulkki2007spatial, kitic2018tramp, pavlidi20153d}. However, it has been observed \cite{evers2019locata,perotin2019crnn} that reverberation introduces significant bias in such DoA estimates, effectively limiting their direct use to anechoic and mildly reverbarent environments. Much research has been dedicated to alleviating the multipath effects, \emph{e.g.} by dereverberating the recorded audio \cite{nakatani2010speech}, by exploiting the Higher Order Ambisonics (HOA) channels \cite{moore2017direction,hafezi2017augmented}, and by estimating the time-frequency bins dominated by the direct path signal \cite{nadiri2014localization,madmoni2018direction}. 

On the other hand, the narrowband reactive sound intensity has been largely ignored, although its physical significance has been recognized \cite{jacobsen1991note,fahy2000foundations}. A recent work \cite{perotin2019crnn} has shown the added value of using this quantity as an input feature for a DoA estimation neural network.

In this article, we propose a wideband representation of velocity vector \cite{daniel2000representation}, which naturally encapsulates both the active and reactive components of complex sound intensity. The derived features are conceptually and computationally simple, yet sufficiently rich to describe the acoustic multipath captured by an Ambisonics microphone. Indeed, as noted in several works \cite{dokmanic2013acoustic, kitic2014hearing, javed2016spherical}, the acoustic reflections are not necessarilly a nuisance, provided that one can identify their parameters properly. 

In the following section, we model the complex sound intensity and the related velocity vector in the presence of strong acoustic reflections, and derive the Time Domain Velocity Vector (TDVV) features. 
In section \ref{sec:algorithm} we illustrate one application of TDVV - the DoA and range estimation of a single, static sound source. Section \ref{sec:experiments} shows the experimental results obtained on the real-life Ambisonics datasets. Section \ref{sec:conclusion} concludes the article. 

\section{Time Domain Velocity Vector}
\label{sec:intensity}

Ambisonics is a spatial audio format based on the spherical harmonic decomposition of a soundfield captured by a spherical microphone array \cite{daniel1998ambisonics,merimaa2006analysis}. In practice, the decomposition is truncated - only a limited number of decomposition coefficients (analogously, audio channels) is retained. The FOA format keeps only four channels: the omnidirectional $w(t)$, and three figure-of-eight channels $x(t)$, $y(t)$ and $z(t)$, aligned with the coordinate axes. 

The channel $w(t)$ designates the acoustic pressure $p(t)$, while the remaining three FOA channels represent its spatial derivatives (all with regards to the same point in space, \emph{i.e.} the microphone position). Since the linearized fluid momentum equation relates the pressure gradient\footnote{Hereafter, we use physics notation ($\vec{\cdot}$) for vector quantities, and boldface letters to denote their estimates (lowercase - vectors, uppercase - matrices).} $\vec{\nabla} p(t)$ and particle velocity $\vec{v}_{\mathrm{p}}(t)$ \cite{merimaa2006analysis}, the $(x,y,z)$ channels are proportional to particle velocity (up to multiplicative constant), \emph{i.e.} $\transp{\left[ \begin{matrix} x(t) \; y(t) \; z(t) \end{matrix} \right]} = \vec{\nabla} p(t) \propto \vec{v}_{\mathrm{p}}(t)$. In the frequency domain\footnote{$S(f) = \mathcal{F}(s(t))$ denotes the Fourier representation of a signal $s(t)$.}, we have $\transp{\left[ \begin{matrix} X(f) \; Y(f) \; Z(f) \end{matrix} \right]} \propto \vec{V}_{\mathrm{p}}(f)$. 
    
The \emph{(frequency-domain) velocity vector} (FDVV) $\vec{V}(f)$ is defined as follows\footnote{Without loss of generality, here we adopt the SID channel ordering with SN3D normalization \cite{daniel2000representation}.} \cite[p.23]{daniel2000representation}:
\begin{equation}\label{eqFDVVcomp}
	\vec{V}(f) = -\frac{1}{\rho c} \frac{\vec{V}_{\mathrm{p}}(f)}{P(f)} = -\frac{j}{k} \frac{\vec{\nabla} P(f)}{P(f)} \simeq \frac{1}{W(f)} \left[ \begin{matrix} X(f) \\ Y(f) \\ Z(f) \end{matrix} \right],
\end{equation}
where $\rho$ is the specific density of air, $c$ is the speed of sound, $k$ is the wavenumber, and $j^2=-1$. 

The FDVV is a normalized version of \emph{complex} sound intensity ${\vec{I}(f) = P(f) \vec{V}_{\mathrm{p}}(f)}^*$ \cite{jacobsen1991note}, with the distinct advantage of being less dependent on the energy $|P(f)|^2$, and thus, the content of the source signal. In the same vein, the active and reactive intesities are proportional to $\Re(\vec{V}(f))$ and $\Im(\vec{V}(f))$ \cite{daniel2000representation}, respectively, where $\Re(\cdot)$ and $\Im(\cdot)$ denote the real and imaginary part of a complex number.

Moreover, FDVV obeys the superposition principle, \emph{i.e.}
\begin{equation}\label{eqSuperposition}
	\vec{V}(f) = \frac{ \sum_{\ing{n}} a_{\ing{n}} \vec{u}_\ing{n} }{\sum_{\ing{n}} a_{\ing{n}}} = \frac{ \vec{u}_\ing{0} + \sum_{\ing{n} \geq 1} \gamma_{\ing{n}} \vec{u}_\ing{n} }{1 + \sum_{\ing{n} \geq 1} \gamma_{\ing{n}}}.
\end{equation} 
Here $a_{\ing{n}}$ and $\vec{u}_\ing{n}$ are the complex magnitude and (unitary) velocity vector of the $\ing{n}$\textsuperscript{th} \emph{plane wave} component, respectively, while $\gamma_{\ing{n}} = a_{\ing{n}} / a_{\ing{0}}$ is the relative gain of the $\ing{n}$\textsuperscript{th} component with respect to the first one ($\ing{n}=0$). 

Assume now a simplified setting, \emph{i.e.} the superposition of the plane wave coming from the source direction $\vec{u}_0$, and the one from the direction $\vec{u}_1$ (\emph{e.g.} representing a dominant reflection, as in Fig.\ref{figSchema}). The FDVV expression \eqref{eqSuperposition} becomes
\begin{equation}\label{eqVV}
	\vec{V}(f) = \frac{\vec{u}_0 + g_1(f) e^{-j2\pi f \tau_1}\vec{u}_1}{1 + g_1(f) e^{-j2\pi f \tau_1}}.
\end{equation}
The complex gain is ${\gamma_1 = g_1(f) e^{-j2\pi f \tau_1}}$, where ${|g_1(f)| < 1}$ represents the frequency-dependent attenuation factor, and $\tau_1$ is the corresponding propagation delay.

Since $|\gamma_1|<1$, the expression \eqref{eqVV} admits the expansion into the Taylor series:
\begin{align}\label{eq_vVel_Taylor1}
    \vec{V}(f) & = \frac{\vec{u}_0 + \gamma_1 \vec{u}_1}{1 + \gamma_1} = (\vec{u}_0 + \gamma_1 \vec{u}_1)\sum_{\ing{k}\geq 0}{(-\gamma_1)^\ing{k}} \nonumber \\
    &= \vec{u}_0 + \sum_{\ing{k}\geq 1}{(-g_1(f))^\ing{k} e^{-2\pi f \ing{k} \tau_1} (\vec{u}_0-\vec{u}_1)}.
\end{align}

Approximating $g_1(f)$ by a frequency-independent constant $g_1$ (as, \emph{e.g.} in \cite{bertin2016joint}), and by applying the inverse Fourier transform $\mathcal{F}^{-1}(\cdot)$ to the series above, we get the TDVV $\vec{v}(t)$ for the one-reflection setting:
\begin{equation}\label{eq_vVel_TD1}
    \vec{v}(t)= \mathcal{F}^{-1}(\vec{V}(f)) = \delta(t) \vec{u}_0 + \sum_{\ing{k}\geq 1}{\delta(t-\ing{k}\tau_1)(-g_1)^{\ing{k}}(\vec{u}_0-\vec{u}_1)}.
\end{equation}
Thus, for $t>0$ the TDVV defines a periodic sequence (with period $\tau_1$), of alternating, exponentially decaying coefficients multiplying the difference vector $\vec{u}_0 - \vec{u}_1$. At $t=0$, TDVV yields the unitary vector colinear with DoA, which is equal to the mean value of $\Re(\vec{V}(f))$ when the frequencies are uniformly distributed. Interestingly, it can be shown that the mode of the probability distribution of $\Re(\vec{V}(f))$ does \emph{not} correspond to its mean. Consequently, peak-picking approaches, such as \cite{kitic2018tramp,pavlidi20153d}, would suffer from a systematic bias in their DoA estimates, even under the one-reflection model.

\begin{figure}
    \includegraphics[width=\columnwidth]{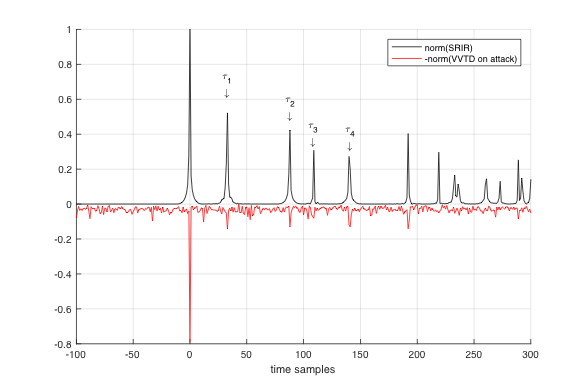}
    \caption{The (mirrored) norm of TDVV against the (centered) norm of Room Impulse Responses of the FOA channels.}\label{figImprint}
\end{figure}

When multiple reflections are allowed, the convergence criterion for the Taylor series becomes much more stringent, \emph{i.e.} it translates into $|\sum_{\ing{n}>0}\gamma_n(f)| < 1, \;\forall f$, which implies $\sum_{n>0}|g_n(f)| < 1$. Then, from \eqref{eqSuperposition} we get
\begin{equation}\label{eq_vVel_TaylorMulti}
    \vec{V}(f) = \left( \vec{u}_\ing{0} + \sum_{\ing{n} \geq 1} \gamma_{\ing{n}} \vec{u}_\ing{n} \right) \sum_{ \ing{k} \geq 0}\left( \sum_{\ing{n}\geq 1} -\gamma_n(f) \right)^{\ing{k}}.
\end{equation}
After the inverse Fourier transform and some manipulations, the TDVV has the following form:
\begin{equation}\label{eq_vVel_TDmulti}
    \vec{v}(t) = \delta(t) \vec{u}_0 + \sum_{\ing{n} \geq 1} \sum_{\ing{k} \geq 1}(-g_{\ing{n}})^{\ing{k}}\delta(t - \ing{k}\tau_{\ing{n}})(\vec{u}_0 - \vec{u}_{\ing{n}}) + \eta.
\end{equation}
In the expression above, we emphasized the contribution of the primary reflections, while the cross terms (interactions among the reflections themselves) are captured by the variable $\eta$. Here again, $\vec{v}(t=0)$ reveals the source DoA, but the ‘‘tail'' is now composed of a linear combination of multiple periodic series, instead of the single one, as in \eqref{eq_vVel_TD1}. 

Unfortunatelly, the convergence condition above is overly restrictive in real acoustics. Empirically, we observed a rather similar behavior in the TDVV of recorded FOA signals (\emph{cf.} Fig.\ref{figImprint}), albeit with some characteristic irregularities. The detailed analysis of the latter phenomena is left for future work.



\section{DOA and range estimation}
\label{sec:algorithm}

\begin{figure}
    \includegraphics[width=\columnwidth]{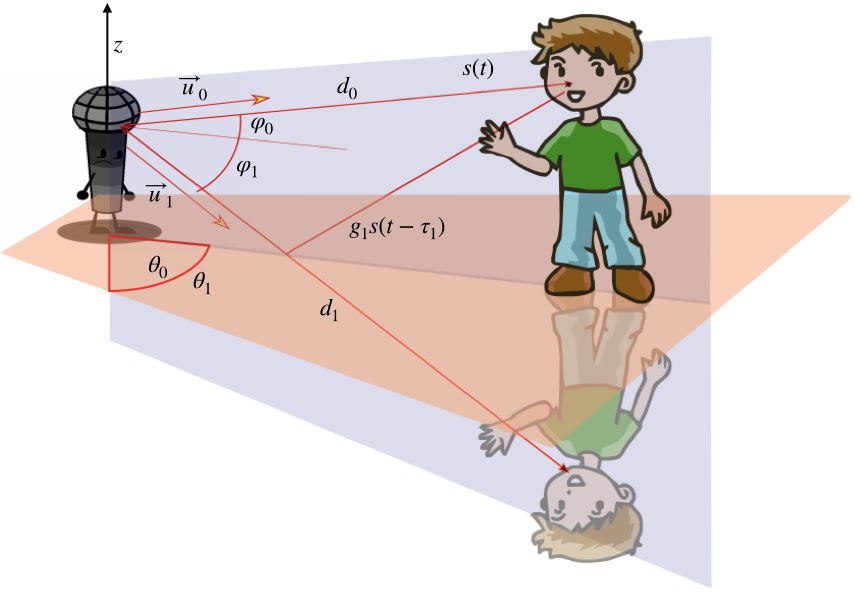}
    \caption{Interference of the direct path and the floor reflection plane waves, by the image-source model. Note that the azimuth angles $\theta_0$ and $\theta_1$ are assumed equal.}\label{figSchema}
\end{figure}

For a showcase application of TDVV, we propose a method to estimate the DoA and range to a sound source, using only the FOA data. The classical methods for range estimation exploit phase differences between channels and the near field propagation model, which is not feasable in our case, since the Ambisonics channels should be (theoretically) in phase. Instead, we exploit the spatial diversity provided by the one-reflection model, schematically presented in Fig.\ref{figSchema}. We assume specular reflections, and a horizontal reflecting surface - \emph{e.g.} a floor, a table, or a ceiling.

First, we discuss the distance estimation. Provided that we have the estimates of the DoA and reflection directions ($\vec{u}_0$ and $\vec{u}_1$), as well as the propagation delay $\tau_1$, simple geometry states that the distances to the source and its image are related as follows:
\begin{equation*}
    d_0 \cos \varphi_0 = d_1 \cos \varphi_1,
\end{equation*}
where $\varphi_0$ and $\varphi_1$ are the elevation angles of $\vec{u}_0$ and $\vec{u}_1$, respectively. Since $d_1 - d_0 = \tau_1 c$, we get the expression for the range $d_0$ as
\begin{equation}\label{eqDistance}
    d_0 = \frac{\tau_1 c}{\frac{\cos \varphi_0}{\cos \varphi_1} - 1}.
\end{equation}
Hence, the goal becomes estimating the unknowns $\vec{u}_0$, $\vec{u}_1$ and $\tau_1$, from the FOA recordings only. While there could be more robust ways to estimate these parameters, we choose a very simple approach that illustrates the usefulness of the TDVV representation. Indeed,
assuming that the one-reflection model described in the previous section holds, the TDVV expression \eqref{eq_vVel_TD1} suffices to estimate all the parameters. 

The signals are processed in the Short Term Fourier Transform (STFT) domain, and estimation is done for each STFT frame independently. The FDVV is computed by the expression \eqref{eqFDVVcomp}, and assembled into the matrix $\iter{\tilde{\mtrx{V}}}{m}$, where $\ing{m}$ denotes the current \emph{frame index}. While the normalization renders the FDVV more invariant to source signal's content, it may also amplify the noise within. Therefore, we run the online noise level estimator \cite{gerkmann2012unbiased}, and suppress the frequency bands having negative (estimated) Signal-to-Noise Ratios (SNRs). Finally, the TDVV is computed from such denoised FDVV, and represented by the matrix $\iter{\mtrx{V}}{\ing{m}} \Rset{3}{T}$. 

According to \eqref{eq_vVel_TD1}, the (normalized) leftmost column of $\iter{\mtrx{V}}{\ing{m}}$ (denoted hereafter by $\vect{u}_0$), corresponds to the DoA vector $\vec{u}_0$. Moreover, due to $|g_1|<1$, for all the remaining $t>0$, the TDVV having the largest magnitude is at the time delay $t = \tau_1$. Hence
\begin{equation}\label{eqTau1}
    \tilde{\tau}_1 = \frac{\ing{j}_{\max}}{f_s}, \quad \ing{j}_{\max} =  \argmax_{\ing{j}>0} \norm{\iter{\mtrx{V}}{\ing{m}}_{:,\ing{j}}}{},
\end{equation}
$f_s$ is the sampling rate and $\iter{\mtrx{V}}{\ing{m}}_{:,\ing{j}}$ is the $\ing{j}$\textsuperscript{th} column of $\iter{\mtrx{V}}{m}$.

Again from \eqref{eq_vVel_TD1}, we deduce that $\iter{\mtrx{V}}{\ing{m}}_{:,\ing{j}_{\max}}$ is colinear with $\vect{u}_0 - \vect{u}_1$, where $\vect{u}_1$ is the (unit norm) estimate of $\vec{u}_1$. Thus, we immediately have
\begin{equation}
    \vect{u}_1 = \vect{u}_0 - \frac{2 \left( \transp{\vect{u}_0} \iter{\mtrx{V}}{\ing{m}}_{:,\ing{j}_{\max}} \right)}{ \norm{\iter{\mtrx{V}}{\ing{m}}_{:,\ing{j}_{\max}}}{}^2 } \iter{\mtrx{V}}{\ing{m}}_{:,\ing{j}_{\max}}.
\end{equation}

In practice, however, even in the noiseless case, the real recordings comprise multiple reflections that may violate the single-reflection model we exploited above. 

One problem are the reflections from other (non-horizontal) surfaces, and the interferences among reflections themselves (\emph{i.e.} the cross terms in the eq. \eqref{eq_vVel_TDmulti}). In order to suppress their contributions in the TDVV imprint, we propose to re-weight the FDVV components accordingly, \emph{before} applying $\mathcal{F}^{-1}(\cdot)$. To determine the weights, we exploit the fact that (\emph{cf.} Fig.\ref{figSchema}) the DoA $\vec{u}_0$ and the desired reflection $\vec{u}_1$ (as well as their difference $\vec{u}_0 - \vec{u}_1$), lie in the subspace orthogonal to the $xy$-plane. By examining the FDVV expression \eqref{eq_vVel_Taylor1}, we see that, for the frequency bands in which the model holds, the imaginary part $\Im(\vec{V}(f))$ is colinear with $\vec{u}_0 - \vec{u}_1$. Using the current estimate $\vect{u}_0$, we design weights $q(f)$ as follows
\begin{equation}
    q(\ing{f}) = \exp \left(-\frac{|\transp{\Im(\iter{\tilde{\mtrx{V}}}{m})}\vect{n}|}{\norm{\Im(\iter{\tilde{\mtrx{V}}}{m})}{}}\right),
\end{equation}
where $\vect{n}$ is the unit normal to the plane defined by $\vect{u}_0$ and the $z$-axis. As a heuristic approach to obtain a slightly better conditioned TDVV, we apply the weighting iteratively, using the new estimate of $\vec{u}_0$ to compute weights each time (usually, after few iterations, the estimate stabilizes).

Another issue is that a frame could be corrupted by the impulse response tail of the preceding ones. Intuitively, the frames corresponding to signal attacks should be sufficiently energetic such that the contribution of previous acoustic propagation becomes negligible. To quantify the attack presence in the interesting frequency bands, we use a (positive) finite difference approximation of the STFT magnitude derivative, summed channel- and frequency-wise, and weighted by $q(\ing{f})$: 
\begin{equation}
    \iter{D}{m} = \max \left(0, \sum\limits_{\ing{f}} q(\ing{f}) \sum\limits_{\ing{i}} \frac{\iter{C(\ing{f})_{\ing{i}}}{m+1} - \iter{C(\ing{f})_{\ing{i}}}{m-1}}{\max\left(\iter{C(\ing{f})_{\ing{i}}}{m+1}, \iter{C(\ing{f})_{\ing{i}}}{m-1} \right) + \epsilon} \right),
\end{equation}
where $\iter{C(\ing{f})_{\ing{i}}}{m}$ denotes the magnitude of the $\ing{i}$\textsuperscript{th} FOA channel at the STFT bin $(\ing{m},\ing{f})$, and $\epsilon>0$ is a small constant. 

Only the frames for which $\iter{D}{m} \geq 0.9 \max_{\ing{m}} \iter{D}{m}$ holds are preserved. Additionally, we filter out the frames for which the distance estimate $\iter{d}{m}_0$ is outside some reasonable limits - in our experiments, between $20$cm and $5$m. This is done in order to avoid heavy outliers, which may arise when the elevation cosine ratio in \eqref{eqDistance} approaches one. Indeed, in the far field setting we have $\varphi_1 \approx \varphi_0$, and the estimation becomes unstable. It is noteworthy that this is usually not an issue with vertical reflecting surfaces, such as walls. However, exploiting these requires a priori knowledge of the azimuthal orientation of the microphone.

\section{Results}
\label{sec:experiments}

The proposed algorithm is tested on two sets of recordings: the Task 1 developement dataset of the recently proposed Acoustic Source LOCAlization and TrAcking (LOCATA) challenge \cite{evers2019locata}, and on our internal dataset used in the publication \cite{perotin2019crnn}. The LOCATA dataset is smaller than the latter, but provides the DoA and range labels, whereas the dataset in \cite{perotin2019crnn} contains the DoA information only. On the other hand, this one perfectly matches the considered use case - a microphone placed on the living room table. As a baseline, we used the TRAMP algorithm, presented in \cite{kitic2018tramp}, which only exploits the active sound intensity.

The sampling frequency is $16$kHz. In order to better capture the signal segments dominated by the direct propagation and horizontal echo, we use very short STFT frames ($\sim 0.005$s), and apply a $95\%$ overlap. For TRAMP, we use the recommended frame length of $0.04$s, and the same overlap size. The frame-level estimates are aggregated by median filtering, hence each algorithm outputs a unique estimate per recording (DoA, or DoA plus range).

Table \ref{tabLOCATA} shows the results obtained on the three recordings comprising the LOCATA Task 1 validation dataset. In all cases, the proposed method outperformes the baseline in terms of the angular error. The average range estimation error is around $20$cm (whereas the average source distance is about $2$m). We conjecture that even greater gains in accuracy could be attained had we used the reflections of vertical surfaces.

\begin{table}
    \centering
    \begin{tabular}{| c | c | c | c |}
        \hline		
        Recording & 1 & 2 & 3 \\	
        \hline
        \hline
        Angular error (TRAMP) & \ang{6.05} & \ang{8.03} & \ang{16.05} \\
        Angular error (proposed) & \ang{3.44} & \ang{3.76} & \ang{9.17} \\
        Range error (proposed) & 32cm & 13cm & 18cm \\
        \hline  
    \end{tabular}
    \caption{Results on the Task 1 developement dataset.}\label{tabLOCATA}
\end{table}

Table \ref{tabStudio} presents the mean and median angular errors for the second dataset. In addition to TRAMP, we added the results, reported in \cite{perotin2019crnn}, of a Convolutional Recurrent Neural Network (CRNN) trained on active and reactive sound intensity features, provided within independent feature channels. Again, we see that the proposed algorithm largely surpasses TRAMP, and even edges out the CRNN model. This indicates that a deep neural network may benefit from the input features given in the TDVV format.

\begin{table}
    \centering
    \begin{tabular}{| r | c | c | c |}
        \hline		
        Angular error & TRAMP & CRNN & Proposed \\	
        \hline
        \hline
        Mean  & \ang{14.45} & \ang{8.1} & \textbf{\ang{6.31}} \\
        Median  & \ang{11.88} & \ang{5.7} & \textbf{\ang{5.45}} \\
        \hline  
    \end{tabular}
    \caption{Results on the dataset used in \cite{perotin2019crnn}.}\label{tabStudio}
    \end{table}




\section{Conclusion}
\label{sec:conclusion}

We presented Time Domain Velocity Vector - a novel spatial representation of Ambisonics soundfields. TDVV is almost independent of the excitation signal, and mainly encodes the propagation signature of the acoustic environment. Next, we demonstrated an application of TDVV to DoA and range estimation of an immobile sound source. Although simple, the proposed algorithm performs favorably to our baseline, strictly by exploiting the properties of the TDVV. 

We postulate that these new features would enable better understanding of the multipath phenomena, from theoretical and practical perspective. For instance, this signal representation would directly benefit the methods for sound source localization and tracking, dereverberation, geometry estimation, beamforming and spatial audio coding. As a consequence, applications spanning from home and professional audio, to virtual and augmented reality, robotics and home/industrial surveillance, are easy to envision.

Future research would focus on better understanding of TDVV in more general settings, \emph{e.g.} in the presence of multiple sources, non-specular reflections, frequency-dependant gains, or when the Taylor series expansion is not viable. From the algorithmic point of view, we are particularly interested in robust estimation methods, either analytic or learning-based. Finally, in this work we considered only the features obtained from the FOA channels. Nevertheless, the extension to HOA is straightforward, and is planned for coming publications.

\bibliographystyle{IEEEbib}
\bibliography{refs}

\begin{thebibliography}{10}

\bibitem{pulkki2007spatial}
V.~Pulkki,
\newblock ``Spatial sound reproduction with directional audio coding,''
\newblock {\em Journal of the Audio Engineering Society}, vol. 55, no. 6, pp.
  503--516, 2007.

\bibitem{daniel1998ambisonics}
J.~Daniel, J.-B. Rault, and J.-D. Polack,
\newblock ``Ambisonics encoding of other audio formats for multiple listening
  conditions,''
\newblock in {\em Audio Engineering Society Convention 105}. Audio Engineering
  Society, 1998.

\bibitem{spors2007extension}
S.~Spors,
\newblock ``Extension of an analytic secondary source selection criterion for
  wave field synthesis,''
\newblock in {\em Audio Engineering Society Convention 123}. Audio Engineering
  Society, 2007.

\bibitem{merimaa2006analysis}
J.~Merimaa,
\newblock {\em Analysis, synthesis, and perception of spatial sound: binaural
  localization modeling and multichannel loudspeaker reproduction},
\newblock Ph.D. thesis, Helsinki University of Technology, 2006.

\bibitem{rossing2007springer}
T.~Rossing~(Ed.),
\newblock {\em Springer handbook of acoustics},
\newblock Springer Science \& Business Media, 2007.

\bibitem{jacobsen1991note}
F.~Jacobsen,
\newblock ``A note on instantaneous and time-averaged active and reactive sound
  intensity,''
\newblock {\em J. of Sound and Vibration}, vol. 147, no. 3, pp. 489--496, 1991.

\bibitem{fahy2000foundations}
F.~Fahy,
\newblock {\em Foundations of engineering acoustics},
\newblock Elsevier, 2000.

\bibitem{daniel2000representation}
J.~Daniel,
\newblock {\em Repr{\'e}sentation de champs acoustiques, application {\`a} la
  transmission et {\`a} la reproduction de sc{\`e}nes sonores complexes dans un
  contexte multim{\'e}dia},
\newblock Ph.D. thesis, University of Paris VI, 2000.

\bibitem{jarrett20103d}
D.~Jarrett, E.~Habets, and P.~Naylor,
\newblock ``{3D source localization in the spherical harmonic domain using a
  pseudointensity vector},''
\newblock in {\em 18th European Signal Processing Conference (EUSIPCO)}. IEEE,
  2010, pp. 442--446.

\bibitem{kitic2018tramp}
S.~Kiti{\'c} and A.~Gu{\'e}rin,
\newblock ``{TRAMP: Tracking by a Real-time AMbisonic-based Particle filter},''
\newblock in {\em LOCATA Challenge Workshop - a satellite event of IWAENC},
  2018.

\bibitem{pavlidi20153d}
D.~Pavlidi, S.~Delikaris-Manias, V.~Pulkki, and A.~Mouchtaris,
\newblock ``{3D localization of multiple sound sources with intensity vector
  estimates in single source zones},''
\newblock in {\em 2015 23rd European Signal Processing Conference (EUSIPCO)}.
  IEEE, 2015, pp. 1556--1560.

\bibitem{evers2019locata}
C.~Evers, H.~Loellmann, H.~Mellmann, A.~Schmidt, H.~Barfuss, P.~Naylor, and
  W.~Kellermann,
\newblock ``{The LOCATA Challenge: Acoustic Source Localization and
  Tracking},''
\newblock {\em IEEE Transactions on Audio, Speech and Language Processing},
  2019.

\bibitem{perotin2019crnn}
L.~Perotin, R.~Serizel, E.~Vincent, and A.~Gu{\'e}rin,
\newblock ``{CRNN-based multiple DoA estimation using acoustic intensity
  features for Ambisonics recordings},''
\newblock {\em IEEE Journal of Selected Topics in Signal Processing}, 2019.

\bibitem{nakatani2010speech}
T.~Nakatani, T.~Yoshioka, K.~Kinoshita, M.~Miyoshi, and B.-H. Juang,
\newblock ``Speech dereverberation based on variance-normalized delayed linear
  prediction,''
\newblock {\em IEEE Transactions on Audio, Speech, and Language Processing},
  vol. 18, no. 7, pp. 1717--1731, 2010.

\bibitem{moore2017direction}
A.~Moore, C.~Evers, and P.~Naylor,
\newblock ``Direction of arrival estimation in the spherical harmonic domain
  using subspace pseudointensity vectors,''
\newblock {\em IEEE/ACM Transactions on Audio, Speech and Language Processing
  (TASLP)}, vol. 25, no. 1, pp. 178--192, 2017.

\bibitem{hafezi2017augmented}
S.~Hafezi, A.~Moore, and P.~Naylor,
\newblock ``Augmented intensity vectors for direction of arrival estimation in
  the spherical harmonic domain,''
\newblock {\em IEEE/ACM Transactions on Audio, Speech and Language Processing
  (TASLP)}, vol. 25, no. 10, pp. 1956--1968, 2017.

\bibitem{nadiri2014localization}
O.~Nadiri and B.~Rafaely,
\newblock ``Localization of multiple speakers under high reverberation using a
  spherical microphone array and the direct-path dominance test,''
\newblock {\em IEEE/ACM Transactions on Audio, Speech, and Language
  Processing}, vol. 22, no. 10, pp. 1494--1505, 2014.

\bibitem{madmoni2018direction}
L.~Madmoni and B.~Rafaely,
\newblock ``Direction of arrival estimation for reverberant speech based on
  enhanced decomposition of the direct sound,''
\newblock {\em IEEE Journal of Selected Topics in Signal Processing}, vol. 13,
  no. 1, pp. 131--142, 2018.

\bibitem{dokmanic2013acoustic}
I.~Dokmani{\'c}, R.~Parhizkar, A.~Walther, Y.~Lu, and M.~Vetterli,
\newblock ``Acoustic echoes reveal room shape,''
\newblock {\em Proceedings of the National Academy of Sciences}, vol. 110, no.
  30, pp. 12186--12191, 2013.

\bibitem{kitic2014hearing}
S.~Kiti{\'c}, N.~Bertin, and R.~Gribonval,
\newblock ``Hearing behind walls: localizing sources in the room next door with
  cosparsity,''
\newblock in {\em 2014 IEEE International Conference on Acoustics, Speech and
  Signal Processing (ICASSP)}. IEEE, 2014, pp. 3087--3091.

\bibitem{javed2016spherical}
H.~Javed, A.~Moore, and P.~Naylor,
\newblock ``Spherical microphone array acoustic rake receivers,''
\newblock in {\em 2016 IEEE International Conference on Acoustics, Speech and
  Signal Processing (ICASSP)}. IEEE, 2016, pp. 111--115.

\bibitem{bertin2016joint}
N.~Bertin, S.~Kiti{\'c}, and R.~Gribonval,
\newblock ``Joint estimation of sound source location and boundary impedance
  with physics-driven cosparse regularization,''
\newblock in {\em 2016 IEEE International Conference on Acoustics, Speech and
  Signal Processing (ICASSP)}. IEEE, 2016, pp. 6340--6344.

\bibitem{gerkmann2012unbiased}
T.~Gerkmann and R.~Hendriks,
\newblock ``Unbiased {MMSE}-based noise power estimation with low complexity
  and low tracking delay,''
\newblock {\em IEEE Transactions on Audio, Speech, and Language Processing},
  vol. 20, no. 4, pp. 1383--1393, 2012.

\end{thebibliography}

\end{document}